\documentclass[a4paper]{jpconf}

\usepackage{graphicx}

\begin{document}
\title{Solution space heterogeneity of the random $K$-satisfiability problem: Theory and simulations}

\author{Haijun Zhou$^{1,2}$}

\address{$^1$Key Laboratory of Frontiers in Theoretical Physics, Institute of Theoretical Physics, Chinese Academy of Sciences,
        Beijing 100190, China}
\address{$^2$Kavli Institute for Theoretical Physics China (KITPC) at the
Chinese Academy of Sciences, Beijing 100190, China}

\ead{zhouhj@itp.ac.cn}

\begin{abstract}
The random $K$-satisfiability ($K$-SAT) problem is an important
problem for studying typical-case complexity of NP-complete
combinatorial satisfaction; it is also a representative model of
finite-connectivity spin-glasses. In this paper we review our recent
efforts on the solution space fine structures of the random $K$-SAT
problem. A heterogeneity transition is predicted to occur in the
solution space as the constraint density $\alpha$ reaches a critical
value $\alpha_{cm}$. This transition marks the emergency of
exponentially many solution communities in the solution space. After
the heterogeneity transition the solution space is still ergodic
until $\alpha$ reaches a larger threshold value $\alpha_d$, at which
the solution communities disconnect from each other to become
different solution clusters (ergodicity-breaking). The existence of
solution communities in the solution space is confirmed by numerical
simulations of solution space random walking, and the effect of
solution space heterogeneity on a stochastic local search algorithm
{\tt SEQSAT}, which performs a random walk of single-spin flips, is
investigated. The relevance of this work to glassy dynamics studies
is briefly mentioned.
\end{abstract}

\section{Introduction}

A random $K$-Satisfiability ($K$-SAT) formula is constructed by
adding $M$ constraints on $N$ variables
\cite{Kirkpatrick-Selman-1994}. Each variable $i$ has a spin state
$\sigma_i= \pm 1$, and each constraint $a$ applies to $K$ different
variables that are randomly chosen from the whole set of variables.
The energy of constraint $a$ is expressed as
\begin{equation}
    \label{eq:energy-sat}
    E_a=\prod\limits_{i\in \partial a} \Biggl[\frac{1-J_a^i \sigma_i}{2} \Biggr] \ ,
\end{equation}
where $\partial a$ denotes the set of variables involved in
constraint $a$ (the size of $\partial a$ is $K$), $J_a^i = \pm 1$ is
the preferred spin state of constraint $a$ on variable $i$. Each of
the $K$ preferred spin values $J_a^i$ of a constraint $a$ are
randomly and independently assigned a value $+1$ or $-1$ with equal
probability. The whole set $\{J_a^i\}$ of preferred spin values are
then fixed, but the actual spin state $\sigma_i$ of each variable
$i$ is allowed to change. The constraint energy $E_a$ is zero if at
least one of the variables $i \in
\partial a$ takes the spin value $\sigma_i = J_a^i$, otherwise
$E_a=1$. Given a random $K$-SAT formula, the task is to construct at
least one spin configuration $\vec{\sigma}\equiv \{ \sigma_1,
\sigma_2, \ldots, \sigma_N \}$ that satisfies all the constraint
(i.e., makes all the constraint energy $E_a$ to be zero), or to
prove that no such solutions (i.e., satisfying spin configurations)
exist.

When $N$ is large, rigorous mathematical proofs (see, e.g., review
article \cite{Achlioptas-2001}) and numerical simulations
\cite{Kirkpatrick-Selman-1994} revealed that whether a random
$K$-SAT formula is satisfiable or not depends on the constraint
density $\alpha$ ($\equiv M/N$). As $\alpha$ increases beyond
certain satisfiability threshold $\alpha_s(K)$, the probability of a
randomly constructed $K$-SAT formula to be satisfiable quickly drops
from being close to unity to being close to zero. Among the whole
ensemble of random $K$-SAT formulas, the satisfiability of those
instances with constraint density $\alpha$ in the vicinity of
$\alpha_s(K)$ is the hardest to determine. This empirical
observation has stimulated a lot of investigations.

The random $K$-SAT problem was intensively studied in the
statistical physics community during the last decade
\cite{Monasson-Zecchina-1996,Mezard-etal-2002,Krzakala-etal-PNAS-2007,Hartmann-Weigt-2005,Mezard-Montanari-2009}.
The number of solutions for the random $K$-SAT problem as a function
of constraint density was calculated in
Ref.~\cite{Monasson-Zecchina-1996} by the replica method of
spin-glass physics. Later the satisfiability threshold $\alpha_s$ as
a function of $K$ was calculated by the first-step
replica-symmetry-breaking (1RSB) energetic cavity method
\cite{Mezard-etal-2002,Mertens-etal-2006}. The evolution of the
solution space of the random $K$-SAT problem with constraint density
$\alpha$ was studied in
Refs.~\cite{Krzakala-etal-PNAS-2007,Montanari-etal-2008,Zhou-2008}
using the 1RSB entropic cavity method
\cite{Mezard-Parisi-2001,Mezard-Montanari-2006}. Before the
satisfiability threshold $\alpha_s(K)$ is reached, the solution
space experiences a clustering transition at $\alpha=\alpha_d(K)$,
where exponentially many solution clusters (Gibbs states) form and
the ergodicity of the solution space is broken. This clustering
transition is followed by a condensation transition at
$\alpha=\alpha_c(K) \geq \alpha_d(K)$, where a sub-exponential
number of solution clusters begin to dominate the solution space.
Within a solution cluster, the spin states of a large fraction of
variables start to be frozen to the same value as $\alpha$ exceeds
certain threshold value $\alpha_f$ that may be different for
different clusters. The freezing transition was investigated in
Refs.~\cite{Montanari-etal-2008,Semerjian-2008,Ardelius-Zdeborova-2008}.
Some of these phase transitions (i.e., the clustering and the
condensation transition) were earlier found to occur in mean-field
$p$-body-interaction spin glasses \cite{Gardner-1985}, with
temperature $T$ (instead of the constraint density)
being the control parameter.

In the present paper, we review our recent efforts on the solution
space fine structures of the random $K$-SAT problem. A heterogeneity
transition is predicted to occur in the solution space as the
constraint density $\alpha$ reaches a critical value
$\alpha_{cm}(K)$ which is smaller than $\alpha_d(K)$
\cite{Zhou-2009-b}. This transition marks the emergency of
exponentially many solution communities in the solution space. For
$\alpha_{cm} < \alpha < \alpha_d$, the heterogeneous solution space
is ergodic; at $\alpha=\alpha_d$ the solution communities will turn
into different solution clusters as an ergodicity-breaking
transition occurs. The existence of solution communities in the
solution space is confirmed by numerical simulations on single
$K$-SAT formulas \cite{Zhou-Ma-2009}, and the effect of solution
space heterogeneity on a stochastic local search algorithm {\tt
SEQSAT}, which performs a random walk of single-spin flips, is
investigated \cite{Zhou-2009}. Beyond the clustering transition
point $\alpha_d$, our numerical simulation results
\cite{Zhou-Ma-2009,Li-Ma-Zhou-2009} suggested that the individual
solution clusters of the solution space also have rich internal
structures.

The replica-symmetric cavity method is used in the next section to
calculate the value of $\alpha_{cm}$ for the onset of solution space
structural heterogeneity. Section~\ref{sec:solution-sampling}
presents the data-clustering results on sampled solutions of single
random $K$-SAT formulas; Section~\ref{sec:seqsat} reports the
simulation results of the stochastic search algorithm {\tt SEQSAT}.
We conclude this work in Sec.~\ref{sec:conclusion} and point out
some possible links with the phenomena of two-step relaxation and
dynamical heterogeneity in supercooled liquids.

\section{Solution space heterogeneity transition}
\label{sec:alpha-cm}

Given a random $K$-SAT formula, the total energy $E(\vec{\sigma})
\equiv \sum_{a} E_a$ is equal to the number of violated constraints
by the spin configuration $\vec{\sigma}$. The whole set
$\mathcal{S}$ of spin configurations with $E(\vec{\sigma})=0$ form
the solution space of this $K$-SAT formula. The similarity between
any two solutions $\vec{\sigma}^1$ and $\vec{\sigma}^2$ of the space
$\mathcal{S}$ can be measured by an overlap parameter defined as
\begin{equation}
    \label{eq:q-def}
    q(\vec{\sigma}^1, \vec{\sigma}^2) = \frac{1}{N}
    \sum\limits_{i=1}^{N} \sigma_i^1 \sigma_i^2 \ .
\end{equation}
To characterize the statistical property of the solution space, we
count the total number of solution-pairs with overlap value $q$ and
denote this number as $\mathcal{N}(q)$. For a random $K$-SAT formula
of size $N\geq 1$ and constraint density $\alpha<\alpha_s(K)$, the
size of $\mathcal{S}$ is exponential in $N$, and $\mathcal{N}(q)$ is
also exponential in $N$. It is helpful to define an entropy density
$s(q)$ as $s(q)\equiv (1/N)\ln[ \mathcal{N}(q)]$.

%
\begin{figure}[t]
    \begin{center}
        \includegraphics[width=0.6\textwidth]{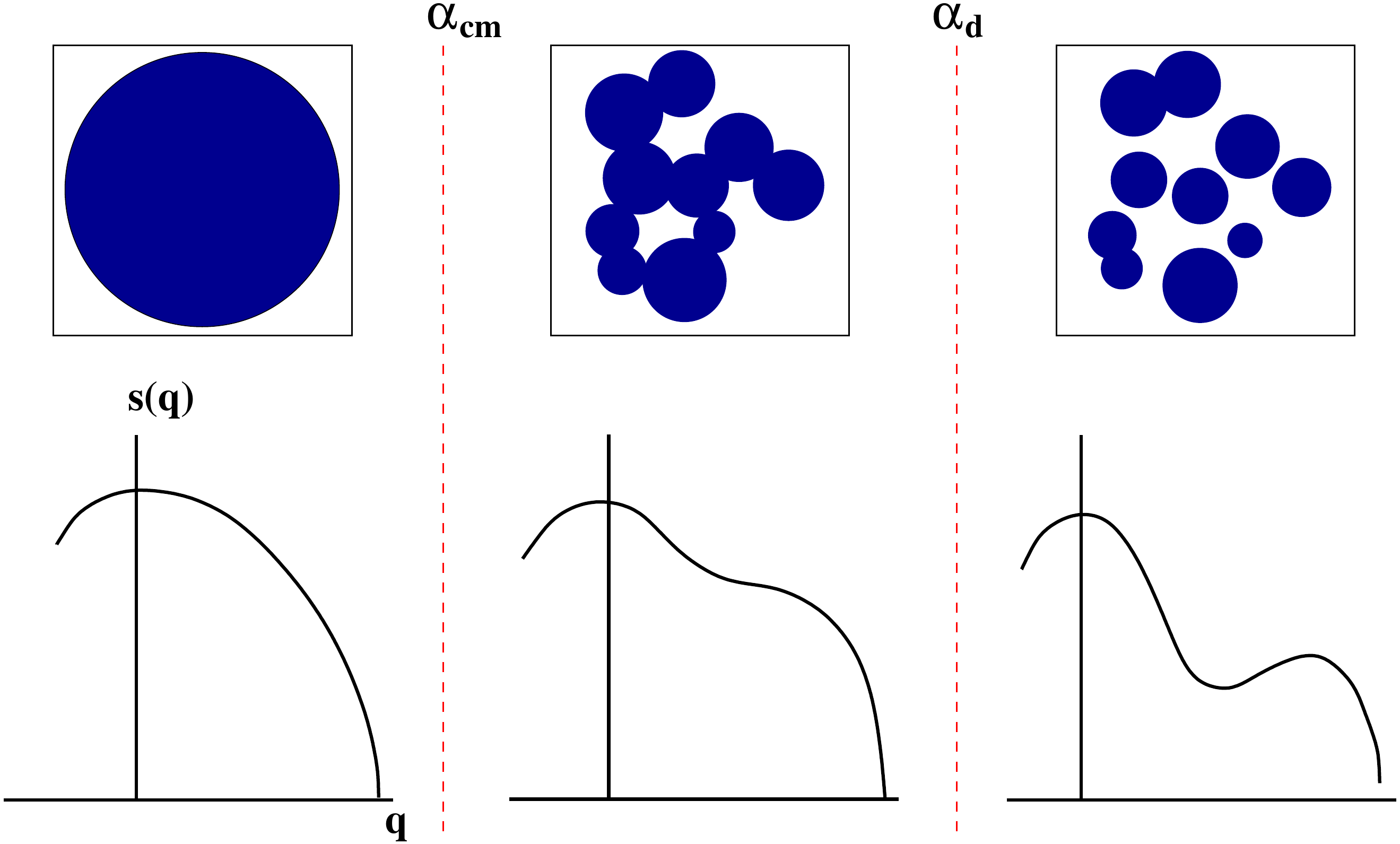}
    \end{center}
    \caption{\label{fig:community}
    Detecting heterogeneity of the solution space by the shape of the entropy
    density function $s(q)$:
    At low constraint density $\alpha$ (left panel), $s(q)$ is a concave
    function of the solution-pair overlap $q$. Solution communities start
    to form as $\alpha$ exceeds a threshold value $\alpha_{cm}$
    (middle panel) and the solution space becomes heterogeneous; this
    makes $s(q)$ to be non-concave. At the clustering transition point
    $\alpha_d$ (right panel), solution communities separate into different
    clusters, and $s(q)$ becomes non-monotonic. As $\alpha$ further
    increases, solution-pairs with intermediate overlap values may cease to
    exist and then $s(q)=-\infty$ for these intermediate $q$ values.
    }
\end{figure}

The solution space $\mathcal{S}$ has a natural graphical
representation (solution graph), in which each solution is denoted
as a node and two nodes are connected by an edge if (and only if)
the corresponding solutions are interchangeable by a single-spin
flip. When $\alpha$ is less than the clustering transition point
$\alpha_d(K)$, this solution graph has only a single connected
component (solution cluster) that contains all (or almost all) of
the solutions \cite{Krzakala-etal-PNAS-2007}. As is well known in
social and biological network studies (see, e.g.,
Refs.~\cite{Wasserman-Faust-1994,Girvan-Newman-2002}), a connected
network component may have rich internal community structures. A
community of a graph refers to a subset of nodes of the graph that
are much more densely inter-connected with each other than with the
remaining nodes of the graph. If communities exist in the solution
graph of a large random $K$-SAT formula, many subsets of solutions
(solution communities) can then be extracted, with the property that
each solution community contains a set of solutions that are much
more similar with each other than with the other solutions.  If such
solution communities are abundant and of considerable sizes, they
will be statistically relevant and will be manifested in the form of
the entropy density function $s(q)$ (see Fig.~\ref{fig:community}).

We define a partition function $Z(x)$ as
\begin{equation}
    \label{eq:Partition_function}
    Z(x) \equiv \sum\limits_{\vec{\sigma}^{1} \in \mathcal{S}}
    \sum\limits_{\vec{\sigma}^{2} \in \mathcal{S}}
    \exp\Bigl( N x q(\vec{\sigma}^{1}, \vec{\sigma}^{2})  \Bigr)
    = \sum\limits_{q} \exp\Bigl[ N \bigl(
    s(q) + x q \bigr) \Bigr] \ ,
\end{equation}
where $x$ is an auxiliary binding field. The free entropy $\Phi(x)$
of the system is defined as $\Phi(x)\equiv \ln Z(x)$. In the
thermodynamic limit of $N\rightarrow \infty$, the free entropy
density $\phi(x)\equiv \Phi(x)/N$ is related to the entropy density
$s(q)$ by
\begin{equation}
    \label{eq:free_entropy_density}
    \phi(x)= \max\limits_{q\in[-1,1]}
    \bigl[s(q)+ x q \bigr] = s\bigl(\overline{q}(x)\bigr)
    +x \overline{q}(x) \ .
\end{equation}
In Eq.~(\ref{eq:free_entropy_density}), $\overline{q}(x)$ is the
mean value of solution-solution overlaps under the binding field
$x$, it is the argument value at which the function $s(q)+x q$
reaches the global maximum.

If the entropy density $s(q)$ is a concave function of $q \in [q_0,
1]$ (Fig.~\ref{fig:community}, left panel), where $q_0$ is the most
probable solution-pair overlap value, then for each $x\geq 0$ there
is only one mean overlap $\overline{q}$, and $\overline{q}(x)$ is a
continuous function of $x$. On the other hand, if $s(q)$ is
non-concave in $q\in [q_0, 1]$ (Fig.~\ref{fig:community}, middle and
right panel), then the value of $\overline{q}(x)$ changes
discontinuously at certain value of $x=x^*>0$. We have exploited
this correspondence between the non-concavity of $s(q)$ and the
discontinuity of $\overline{q}(x)$ to determine the threshold
constraint density $\alpha_{cm}$ at which $s(q)$ starts to be
non-concave \cite{Zhou-2009-b}. We regard $\alpha=\alpha_{cm}$ as
the point at which the solution space of the random $K$-SAT problem
transits into structural heterogeneity. This is because, as
schematically shown in Fig.~\ref{fig:community}, at $\alpha_{cm}$ it
starts to make sense to distinguish between intra-community overlap
values and inter-community overlap values. As shown in
Sec.~\ref{sec:solution-sampling}, many solution communities can be
identified in a heterogeneous solution space \cite{Zhou-Ma-2009}.
Each solution community contains a set of solutions which are more
similar with each other than with the solutions of other
communities. These differences of intra- and inter-community overlap
values and the relative sparseness of solutions at the boundaries
between solution communities cause the non-concavity of $s(q)$.

For the random $K$-SAT problem, we use the replica-symmetric cavity
method of statistical mechanics \cite{Mezard-Parisi-2001} to
calculate the mean overlap values $\overline{q}(x)$ at each value of
$\alpha < \alpha_d(K)$. As the partition function
Eq.~(\ref{eq:Partition_function}) is a summation over pairs of
solutions $(\vec{\sigma}^{1}, \vec{\sigma}^{2})$, the state of each
vertex is a pair of spins $(\sigma, \sigma^\prime)$. Details of this
calculation can be found in Ref.~\cite{Zhou-2009-b} and here we cite
the main results.

\begin{figure}[t]
    \begin{center}
    \includegraphics[width=0.498\textwidth]{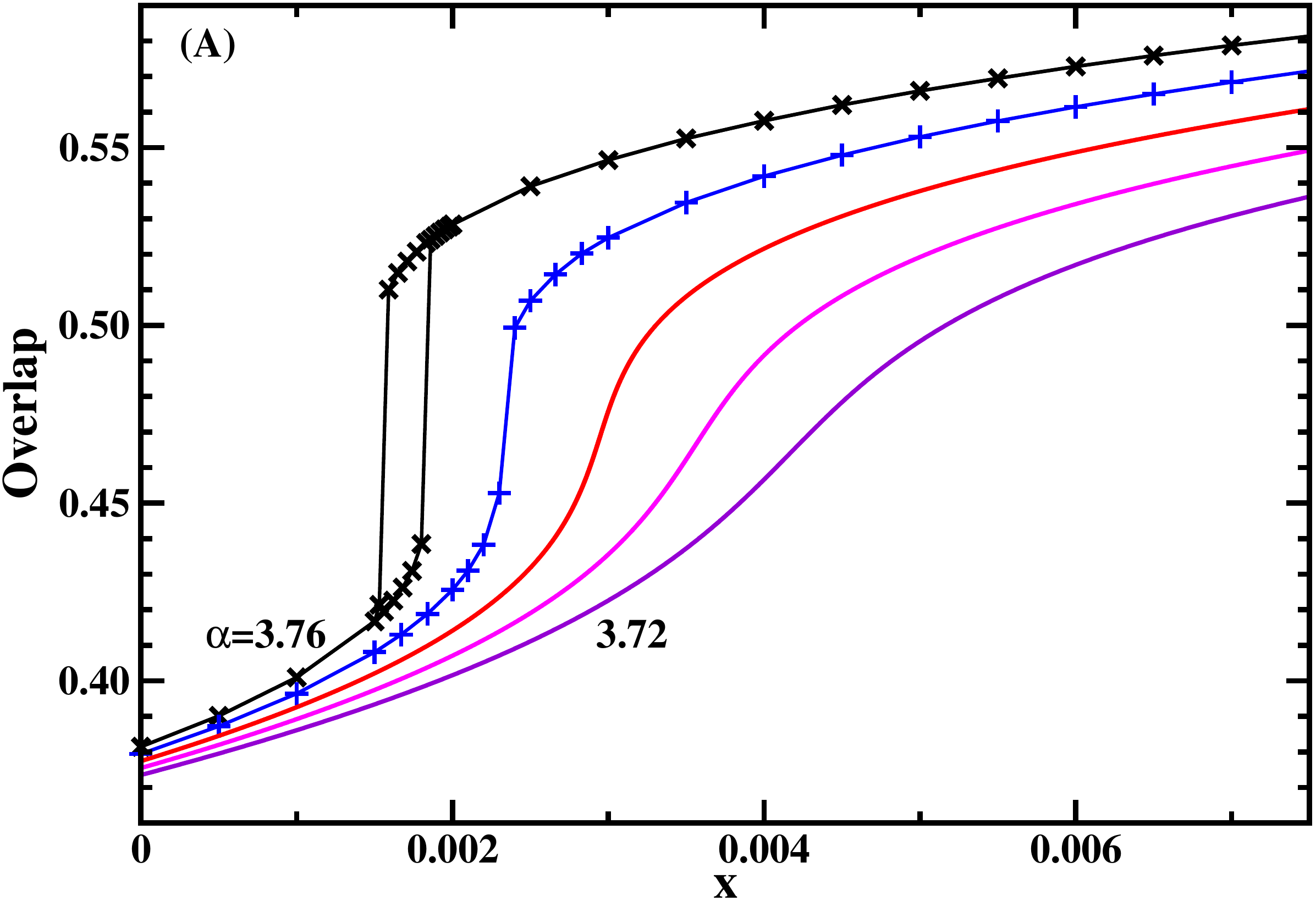}
    \includegraphics[width=0.49\textwidth]{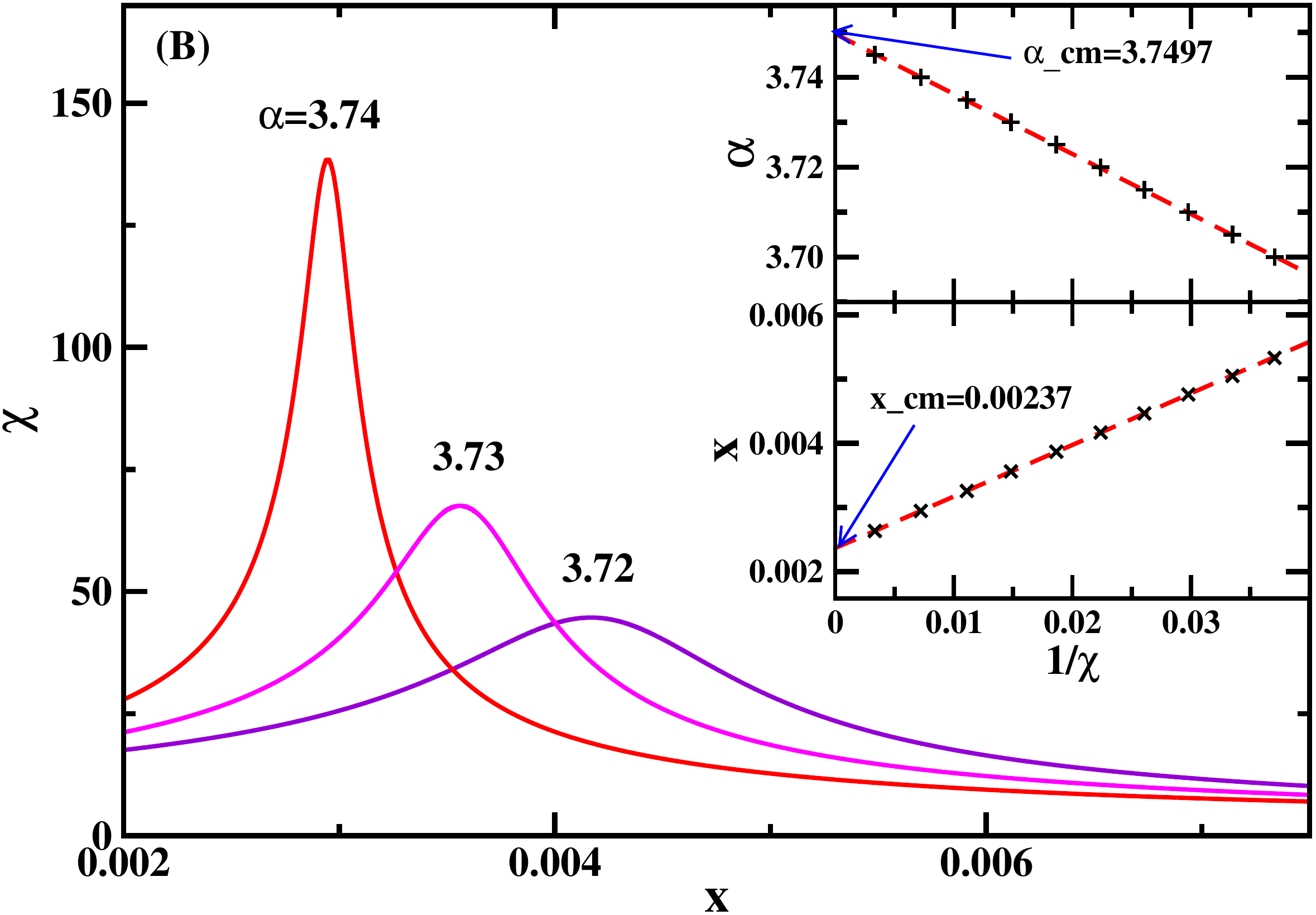}
    \end{center}
    \caption{
    \label{fig:3sat-population}
    The mean overlap $\overline{q}(x)$ and the
    susceptibility $\chi(x)$ at
    constraint density value $\alpha$ for the random
    $3$-SAT problem. In (A) $\alpha$ increases from
    $3.72$ to $3.76$ (right to left) with step size $0.01$.
    The insets of (B) demonstrate that the peak value
    of $\chi$ diverges inverse linearly with $\alpha$ and $x$ as
    the critical point $(\alpha_{cm}=3.7497, x_{cm}=0.0024)$
    is approached.
    }
\end{figure}

Figure~\ref{fig:3sat-population}A shows, for the random $3$-SAT
problem, the form of the function $\overline{q}(x)$ at several
different $\alpha$ values. When $\alpha < \alpha_{cm}(3) \approx
3.75$, the mean overlap $\overline{q}$ increases with the binding
field $x$ smoothly, indicating that the entropy density function
$s(q)$ is concave in shape (Fig.~\ref{fig:community}, left panel).
The solution space $\mathcal{S}$ of the random $3$-SAT problem is
then regarded as homogeneous. When $\alpha> \alpha_{cm}(3)$, there
is a hysteresis loop in the $\overline{q}(x)$ curve as the binding
field $x$ increases and then decreases around certain threshold
value $x^*$. This behavior is typical of a first-order
phase-transition. At $x>x^*$ the partition function $Z(x)$ is
contributed mainly by intra-community solution pairs, while at
$x<x^*$ it is contributed mainly by inter-community solution pairs.

To determine precisely the critical value $\alpha_{cm}$, we
investigate the overlap susceptibility $\chi \equiv {\rm d}
\overline{q}(x) /{\rm d} x$, which is a measures of the overlap
fluctuations,
\begin{equation}
    \chi(x) =
\frac{1}{N}\sum\limits_{i=1}^{N}\sum\limits_{j=1}^{N} \Bigl[
\langle \sigma_i^{1}\sigma_i^{2} \sigma_j^{1} \sigma_j^{2} \rangle
    - \langle \sigma_i^{1}\sigma_i^{2} \rangle
    \langle \sigma_j^{1} \sigma_j^{2}\rangle\Bigr] \ ,
\end{equation}
where $\langle \ldots \rangle$ means averaging over solution-pairs
under the binding field $x$. From the divergence of the peak value
of $\chi(x)$ as shown in Fig.~\ref{fig:3sat-population}B, we obtain
that $\alpha_{cm}(3)=3.7497$ for the random $3$-SAT problem. This
value is much below the value of $\alpha_d(3)=3.87$
\cite{Krzakala-etal-PNAS-2007}.

For the random $4$-SAT problem, we find that
$\alpha_{cm}(4)=8.4746$. This value is again much below the
clustering transition point $\alpha_d(4)=9.38$
\cite{Krzakala-etal-PNAS-2007}. The difference
$\alpha_{d}(K)-\alpha_{cm}(K)$ appears to be an increasing function
of $K$.

\section{Solution graph random walking}
\label{sec:solution-sampling}

For single random $K$-SAT formulas, the solution space structural
heterogeneity can also be detected by performing a long-time random
walking in the corresponding solution graphs
\cite{Zhou-2003-b,Zhou-2003-c,Li-Ma-Zhou-2009,Zhou-Ma-2009}. The
Hamming distance $d(\vec{\sigma}^1, \vec{\sigma}^2)$ between two
solutions $\vec{\sigma}^1$ and $\vec{\sigma}^2$ of the solution
space is defined as
\begin{equation}
 d(\vec{\sigma}^1, \vec{\sigma}^2) = \sum\limits_{i=1}^{N} \Bigl[
  1- \delta(\sigma_i^1, \sigma_i^2) \Bigr]
\end{equation}
where $\delta(\sigma, \sigma^\prime)=1$ if $\sigma=\sigma^\prime$
and $\delta(\sigma, \sigma^\prime)=0$ if $\sigma=-\sigma^\prime$.
This distance counts the number of different spins between the two
solutions. The Hamming distance $d(\vec{\sigma}^1, \vec{\sigma}^2)$
is related to the overlap $q(\vec{\sigma}^1, \vec{\sigma}^2)$
[Eq.~(\ref{eq:q-def})] through
\begin{equation}
    \label{eq:overlap}
    q(\vec{\sigma}^1, \vec{\sigma}^2)
    = 1- \frac{2 d( \vec{\sigma}^1, \vec{\sigma}^2)}{N} \ .
\end{equation}
In the solution graph of a satisfiable random $K$-SAT formula, a
solution $\vec{\sigma}$ is linked to $k_{\vec{\sigma}}$ other
solutions, all of which have unit Hamming distance with
$\vec{\sigma}$. It was empirically found that the degrees
$k_{\vec{\sigma}}$ of the solutions are narrowly distributed with a
mean much less than $N$ \cite{Zhou-Ma-2009}. The solutions can
therefore be regarded as equally important in terms of connectivity.
However, the connection pattern of the solution graph can be highly
heterogeneous. Even when the whole graph is connected, solutions may
still form different communities such that the edge density of a
community is much larger than that of the whole graph. The communities
may even further organize into super-communities.

Consider two solutions $\vec{\sigma}^1$ and $\vec{\sigma}^2$ of the
solution graph. The shortest-path length $l(\vec{\sigma}^1,
\vec{\sigma}^2)$ between these two solutions in the solution graph
satisfies the inequality $l(\vec{\sigma}^1, \vec{\sigma}^2)\geq
d(\vec{\sigma}^1, \vec{\sigma}^2)$. If $\vec{\sigma}^1$ and
$\vec{\sigma}^2$ belong to the same solution community,
$l(\vec{\sigma}^1, \vec{\sigma}^2)$ may be equal to or just slightly
greater than the Hamming distance $d(\vec{\sigma}^1,
\vec{\sigma}^2)$. On the other hand, if $\vec{\sigma}^1$ and
$\vec{\sigma}^2$ belong to two different solution communities, an
extensive spin rearrangement may have to be made to change from
$\vec{\sigma}^1$ to $\vec{\sigma}^2$ by single-spin flips, and then
$l(\vec{\sigma}^1, \vec{\sigma}^2)$ is much greater than
$d(\vec{\sigma}^1, \vec{\sigma}^2)$. If solutions are sampled by a
random walking process at equal time interval $\Delta t$, the
sampled solutions should contain useful information about the
community structure of the solution space, with a resolution level
depending on $\Delta t$. This is because that, when the edges are
followed randomly by a random walker, the walker will be trapped in
different communities most of the time, and the sampled solutions
will then form different similarity groups.

\begin{figure}[t]
    \begin{center}
    \includegraphics[width=0.99\linewidth]{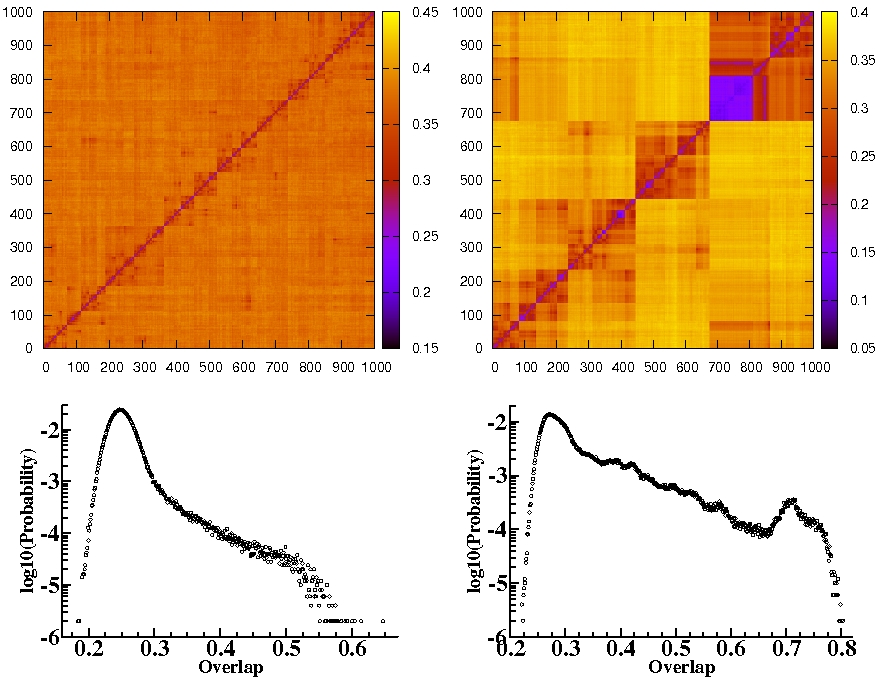}
    \end{center}
    \caption{\label{fig:transition}
    The matrix of Hamming distances of $1000$ sampled solutions for
    a random $4$-SAT formula of $N=20,000$ variables, and the
    corresponding overlap distribution of these sampled solutions.
    The solutions were obtained by an unbiased random walking process
     (solution sampling began after running the random walking process
     for $3\times 10^7 \times N$ steps starting from an initial
     solution, and two consecutively sampled solutions are separated by
     $\Delta t=50,000 \times N$ random walking steps).
    The left panel corresponds to $\alpha=9.10$ and the right panel to
    $\alpha=9.22$. In the Hamming distance matrices, the solutions
    are ordered according to the results of the minimal-variance
    clustering analysis.
    }
\end{figure}

Starting from an initial solution $\vec{\sigma}^0$, a sequences of
solutions $(\vec{\sigma}^0, \vec{\sigma}^{1}, \ldots,
\vec{\sigma}^{l}, \vec{\sigma}^{l+1}, \ldots)$ are generated by
solution graph random walking. Two different random walking
processes are used in the simulation. In the {\em unbiased} random
walking process, the solution $\vec{\sigma}^{l+1}$ is a nearest
neighbor of solution $\vec{\sigma}^{l}$ for $l=0,1,\ldots$
\cite{Zhou-Ma-2009}. Under this simple dynamics, if the generated
solution sequence is infinitely long, each solution $\vec{\sigma}$
in a connected component of the solution graph will appear in the
sequence with frequency proportional to its connectivity
$k_{\vec{\sigma}}$. On the other hand, in the {\em uniform} random
walking process, with probability $k_{\vec{\sigma}^l}/N$, the
solution $\vec{\sigma}^{l+1}$ ($l=0,1,\ldots$) is a nearest neighbor
of solution $\vec{\sigma}^l$, and with the remaining probability
$1-k_{\vec{\sigma}^l}/N$, solution $\vec{\sigma}^{l+1}$ is identical
to $\vec{\sigma}^l$. Under this later dynamical rule, each solution
in a connected component of the solution graph will have the same
frequency to appear in the generated solution sequence. Both these
two random walking processes were used to sample solutions, and the
clustering analysis performed on these two sets of data gave
qualitatively the same results \cite{Zhou-Ma-2009}. The simulation
results shown in Fig.~\ref{fig:transition} were obtained by the
unbiased random walking process.

A set of solutions are sampled (with equal time interval $\Delta t$)
from the generated long sequence of solutions for clustering
analysis. (To avoid strong dependence on the input solution
$\vec{\sigma}^0$, the initial part of the solution sequence was not
used for sampling.) By calculating the overlap values between the
sampled solutions, we obtain an overlap histogram as shown in
Fig.~\ref{fig:transition}. A hierarchical minimum-variance
clustering analysis \cite{Jain-Dubes-1988} (see also
Ref.~\cite{Barthel-Hartmann-2004}) is performed on the sampled
solutions. Initially each solution is regarded as a group, and the
distance between two groups is the Hamming distance. At each step of
the clustering, two groups $C_a$ and $C_b$ that have the smallest
distance are merged into a single group $C_c$. The distance between
the merged group $C_c$ and another group $C_d$ is calculated by
\begin{equation}
    \label{eq:clusterdistance}
    d(C_c, C_d) = \frac{(|C_a| + |C_d|) d(C_a, C_d)
    + (|C_b|+|C_d|) d(C_b, C_d) - |C_d| d(C_a, C_b)}{|C_c|+ |C_d|} \ ,
\end{equation}
where $|C|$ is the number of solutions in group $C$. After the
sampled solutions are listed in the order as reported by the
minimal-variance clustering algorithm, the matrix of Hamming
distances between these solutions are represented in a graphical
form as shown in Fig.~\ref{fig:transition} (upper row).

For a random $4$-SAT formula with $N=20,000$ variables, the overlap
histogram and the Hamming distance matrix of $1000$ sampled
solutions are shown in Fig.~\ref{fig:transition} for two different
constraint density values $\alpha=9.10$ and $\alpha=9.22$ in the
ergodic phase. At $\alpha=9.10$, the Hamming distance matrix has a
weak signature of the existence of many solution communities, and
the overlap histogram is slightly non-concave. At $\alpha=9.22$, the
Hamming distance matrix has a very clear block structure and the
overlap histogram is non-monotonic. These observations suggest that
the explored solution spaces are heterogeneous both at $\alpha=9.10$
and $\alpha=9.22$; the entropic trapping effect of the solution
communities becomes much stronger as $\alpha$ increases from $9.10$
to $9.22$. The simulation results are consistent with the analytical
results of the preceding section, they are also in agreement with
the expectation that, as $\alpha_{d}(4)$ is approached from below, a
random walker becomes more and more trapped in a single solution
community and finally becomes impossible to escape from this single
community at $\alpha \geq \alpha_{d}(4)$.

We have obtained similar simulation results on single random $3$-SAT
formulas \cite{Zhou-Ma-2009}. The solution space random walking
simulations confirm that the solution space of the random $K$-SAT
problem is already very heterogeneous before the clustering
transition point $\alpha_d(K)$ is reached. The simulation results
reported in Ref.~\cite{Zhou-Ma-2009} also suggested that, at
$\alpha> \alpha_d(K)$, the single solution clusters of the solution
space of the random $K$-SAT problem are themselves quite
heterogeneous in internal structure, which may be the reason
underlying the non-convergence of the belief-propagation iteration
process within a single solution cluster \cite{Li-Ma-Zhou-2009}.

\section{Glassy behavior of a stochastic search algorithm}
\label{sec:seqsat}

After a slight change, the solution graph random walking process of
the preceding section was turned into a stochastic local search
algorithm \cite{Zhou-2009}. This algorithm, referred to as {\tt
SEQSAT}, satisfies sequentially the constraints of a random $K$-SAT
formula $F$. We denote by $F_m$ the sub-formula containing the first
$m$ constraints of $F$. Suppose a configuration $\vec{\sigma}(t_m)$
that satisfies $F_m$ is reached at time $t_m$. The $(m+1)$-th
constraint of formula $F$ is added to $F_m$ to obtain the enlarged
sub-formula $F_{m+1}$. Then, starting from $\vec{\sigma}(t_m)$, an
unbiased random walk process is running on the solution graph of
$F_m$ until a spin configuration $\vec{\sigma}(t_{m+1})$ that
satisfies $F_{m+1}$ is first reached at time $t_{m+1}=t_m+ \Delta
t_{m}$ after $\Delta t_{m}\times N$ single-spin flips, $N$ being the
total number of variables in formula $F$. The waiting time of
satisfying the $(m+1)$-th constraint is $\Delta t_{m}$ (this time is
zero if $\vec{\sigma}(t_m)$ already satisfies $F_{m+1}$). Starting
from a completely random initial spin configuration
$\vec{\sigma}(0)$ and an empty sub-formula $F_0$, every constraint
of the formula $F$ are satisfied by {\tt SEQSAT} in this way. Notice
that if a constraint was satisfied by {\tt SEQSAT}, it remains to be
satisfied as new constraints are added (energy barrier crossing is
not allowed).

\begin{figure}[t]
\begin{center}
\includegraphics[width=0.45\textwidth]{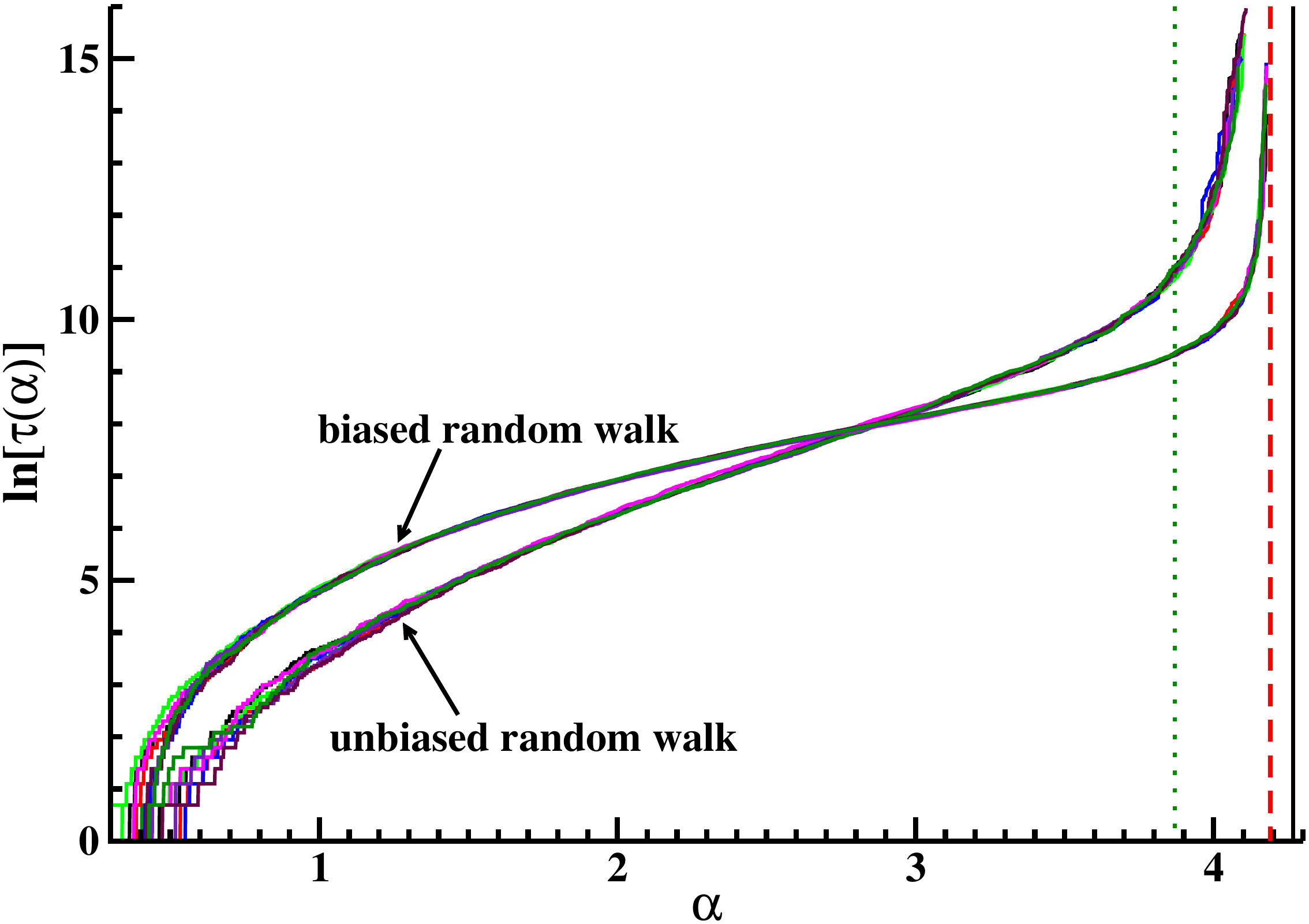}
\hspace*{1.0cm}
\includegraphics[width=0.45\textwidth]{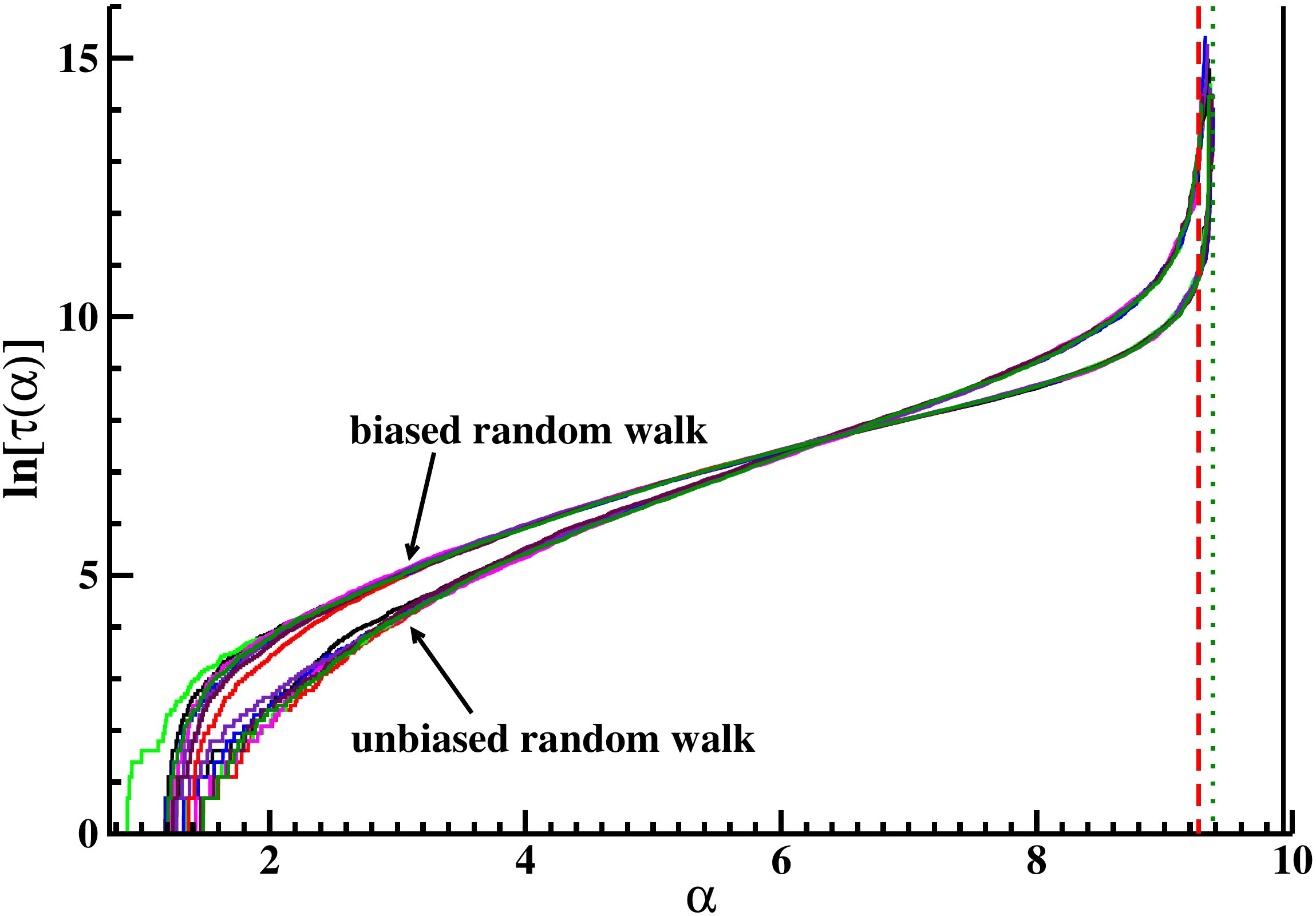}
\end{center}
\caption{\label{fig:34SAT}
    The total search time $\tau(\alpha)$ needed to satisfy the first $\alpha \times N$
    constraints of a random $K$-SAT formula with $N=100,000$ variables
    ($K=3$, left panel; $K=4$, right panel).
    The performances of an unbiased random-walk search process
    and a biased random-walk search process are
    compared. The red dashed lines correspond to the jamming transition points
    $\alpha_j(K)$ as predicted by a long-range-frustration mean-field theory
    \cite{Zhou-2005b,Zhou-2009}, the green dotted lines correspond to
    the clustering transition points $\alpha_d(K)$ \cite{Krzakala-etal-PNAS-2007},
    and the black solid lines correspond to the satisfiability threshold
    value $\alpha_s(K)$ \cite{Mertens-etal-2006}.
    }
\end{figure}

Figure~\ref{fig:34SAT} shows the simulation results of {\tt SEQSAT}
on a random $3$-SAT formula and a random $4$-SAT formula, $N=10^5$.
When the constraint density $\alpha$ of the satisfied sub-formula
$F_m$ is low, the waiting time needed to satisfy a constraint is
very close to zero. As the heterogeneity transition point
$\alpha_{cm}(K)$ is reached, however, the waiting time increases
quickly and it starts to take more than $100 \times N$ single-spin
flips to satisfy a constraint. The search process becomes more and
more slow as $\alpha$ further increases. As $\alpha$ approaches
another threshold value $\alpha_j(K)$ the waiting time is so long
(exceeding $10^6\times N$ single-spin flips) that ${\tt SEQSAT}$
essentially stops to satisfy a newly added constraint. The parameter
$\alpha_{j}(K)$ is regarded as the jamming point of the random walk
search algorithm. In the range of $\alpha_{cm}(K) < \alpha <
\alpha_{j}(K)$ the {\tt SEQSAT} is performing an increasingly
viscous diffusion in the solution space of the random $K$-SAT
formula. The simulation results of Fig.~\ref{fig:34SAT} and those
reported in Ref.~\cite{Zhou-2009} for single random $5$-SAT and
$6$-SAT formulas clearly demonstrate that, the solution space
heterogeneity transition at $\alpha_{cm}(K)$ has significant
dynamical consequences for stochastic local search processes.

For the random $3$-SAT problem, the jamming point $\alpha_j(3)$ is
significantly larger than the clustering transition point
$\alpha_d(3)$. At $\alpha_d(3)$ the solution space of a large random
$3$-SAT formula is dominated by a sub-exponential number of solution
clusters \cite{Krzakala-etal-PNAS-2007}. A dominating solution
cluster also has internal community structures \cite{Zhou-Ma-2009}.
As $\alpha$ further increases, a dominating cluster shrinks in size,
and it break into many sub-clusters. Figure~\ref{fig:34SAT} (left
panel) indicates that, the random walker of {\tt SEQSAT} is residing
on one of the dominating clusters at $\alpha\approx \alpha_d(3)$ and
it continue to be residing on one of the dominating sub-clusters of
the visited cluster as $\alpha$ increases. If the spin values of a
large fraction of variables become frozen in the residing solution
cluster, a jamming transition then occurs. The value of
$\alpha_j(3)$ is predicted to be $4.1897$ by a long-range
frustration mean-field theory \cite{Zhou-2005b,Zhou-2009}, in
agreement with the simulation results.

For the random $K$-SAT problem with $K\geq 4$, the simulation
results shown in Fig.~\ref{fig:34SAT} (right panel) and in
Ref.~\cite{Zhou-2009} suggest that the jamming transition point
$\alpha_j(K)$ is identical to or very close to the clustering
transition point $\alpha_d(K)$. We notice that for the random
$K$-SAT problem with $K\geq 4$, at $\alpha=\alpha_d(K)$, the union
of an exponential number of small solution clusters is contributing
predominantly to the solution space \cite{Krzakala-etal-PNAS-2007}.
It is expected that, as $\alpha$ approaches $\alpha_d(K)$ from
below, the ergodic solution space of a large $K$-SAT formula is
dominated by an exponential number of small solution communities.
The solutions reached by {\tt SEQSAT} probably are residing on one
of these small communities. As $\alpha\rightarrow \alpha_d(K)$, each
of these small (but statistically relevant) solution communities
probably contains a large fraction of variables that are almost
frozen.

Figure~\ref{fig:34SAT} also shows the performances of a biased
random walk search algorithm. The biased random walk process differs
from the unbiased random walk process in that, in each single-spin
flip, a variable that is flippable but not yet being flipped is
flipped with priority \cite{Zhou-2009}. This biased random walk {\tt
SEQSAT} algorithm was implemented with the hope of escaping from a
solution community more quickly. For the random $3$-SAT and $4$-SAT
problem, the biased {\tt SEQSAT} algorithm indeed is more efficient
than the unbiased algorithm. But for the random $K$-SAT problem with
$K\geq 5$, the biased algorithm diverges earlier than the unbiased
algorithm \cite{Zhou-2009}.

\section{Conclusion and discussions}
\label{sec:conclusion}

The solution space structure of the random $K$-SAT problem evolves
with the constraint density $\alpha$. Several qualitative
transitions occurs in the solution space as $\alpha$ becomes
relatively large. We demonstrated, both theoretically and by
computer simulations, that the first qualitative structural
transition is the heterogeneity transition at
$\alpha=\alpha_{cm}(K)$, where exponentially many solution
communities start to form in the (still ergodic) solution space.
The dynamic
behavior of a stochastic search algorithm was investigated. This
simple algorithm {\tt SEQSAT} constructs satisfying spin
configurations for a single $K$-SAT formula by performing a random
walk of single-spin flips. Due to the entropic trapping effect of
solution communities, the solution space random walking process
starts to be very viscous as $\alpha$ goes beyond $\alpha_{cm}(K)$.
{\tt SEQSAT} is able to find solutions for a random $K$-SAT formula
with constraint density less than a threshold value $\alpha_j(K)$.
For $K=3$, the jamming point $\alpha_j(3)\approx 4.19$ is larger
than the solution space clustering transition point $\alpha_d(3)$.
But for $K\geq 4$, it appears that $\alpha_j(K)$ is very close to
$\alpha_d(K)$.

When the constraint density $\alpha$ of a large random $K$-SAT
formula is in the region of $\alpha_{cm}(K) < \alpha <
\alpha_{d}(K)$, the solution space of the formula is heterogeneous
but ergodic. If a random walking process is running on the solution
space starting from an initial solution, a two-step relaxation
behavior can be observed \cite{Zhou-Ma-2009}, corresponding to a
quick slipping into a solution community, a relatively long
wandering within  this community, and finally the viscous diffusing
in the whole solution space. Similar two-step relaxation behaviors
were observed in glassy dynamics studies of supercooled liquids
(see, e.g., the review article \cite{Cavagna-2009}) both in the
non-activated dynamics region and in the activated dynamics region.

The existence of solution communities in the solution space may lead
to a phenomenon that is similar to the dynamical heterogeneity of
supercooled liquids (see, e.g., review articles
\cite{Ediger-2000,Glotzer-2000}). For a random walking process
on a heterogeneous solution space, if the observation time is less
than the typical relaxation time of escaping from the solution
communities, one may find that the spin values of a large fraction
of variables change only very infrequently among the visited
solutions, while the spin values of the remaining variables are
flipped much more frequently. The variables of the $K$-SAT formula
then divides into an active group and an inactive group in this
observation time window. One may further observe that the active
variables are clustered into many distantly separated sub-groups,
each of which containing variables that are very close to each other
\cite{note4}. Numerical simulations need to be performed to
investigate in more detail the solution space random walking
processes.

The constraint density $\alpha$ of the random $K$-SAT problem
corresponds to the particle density $\rho$ in supercooled liquid. In
supercooled liquids, another important control parameter is the
temperature $T$. We can also introduce a positive temperature $T$ to
the random $K$-SAT problem and investigate how the configuration
space evolves with $T$. It is anticipated that, as $T$ is lowered to
a threshold value $T_{cm}(K, \alpha)$ the configuration space will
also experience a heterogeneity transition \cite{Zhou-2009-b}.
Detailed theoretical and simulation results will be reported in a
later work. For the fully-connected $p$-spin spherical model of
glasses, it has already been shown
\cite{Franz-Parisi-1995,Franz-Parisi-1997} that a quantity called the
Franz-Parisi potential has a change in its concavity property
as $T$ is lowered (see \cite{Zdeborova-Krzakala-2010} for more
recent extended studies). Another related problem, 
the weight space property of the Ising perceptron,
was studied in \cite{Obuchi-Kabashima-2009}. For this fully
connected model, a change of concavity property was also predicted for
its characteristic function.

The solution space heterogeneity transition also exists in the
random $K$-XORSAT problem \cite{Zhou-2009}. It may be a general
feature of the configuration spaces of spin glass models on
finite-connectivity random graphs.

\ack

The author thanks Kang Li, Hui Ma and Ying Zeng for collaborations.
This work was partially supported by the National Science Foundation
of China (Grant numbers 10774150 and 10834014) and the China
973-Program (Grant number 2007CB935903).

\section*{References}


\begin{thebibliography}{10}
\expandafter\ifx\csname url\endcsname\relax
  \def\url#1{{\tt #1}}\fi
\expandafter\ifx\csname urlprefix\endcsname\relax\def\urlprefix{URL
}\fi \providecommand{\eprint}[2][]{\url{#2}}

\bibitem{Kirkpatrick-Selman-1994}
Kirkpatrick S and Selman B 1994 {\em Science\/} {\bf 264} 1297--1301

\bibitem{Achlioptas-2001}
Achlioptas D 2001 {\em Theor. Comput. Sci.\/} {\bf 265} 159--185

\bibitem{Monasson-Zecchina-1996}
Monasson R and Zecchina R 1996 {\em Phys. Rev. Lett.\/} {\bf 76}
3881--3885

\bibitem{Mezard-etal-2002}
M{\'{e}}zard M, Parisi G and Zecchina R 2002 {\em Science\/} {\bf
297} 812--815

\bibitem{Krzakala-etal-PNAS-2007}
Krzakala F, Montanari A, {Ricci-Tersenghi} F, Semerjian G and
Zdeborova L 2007
  {\em Proc. Natl. Acad. Sci. USA\/} {\bf 104} 10318--10323

\bibitem{Hartmann-Weigt-2005}
Hartmann A~K and Weigt W 2005 {\em Phase Transitions in
Combinatorial
  Optimization Problems\/} (Weinheim, Germany: Wiley-VCH)

\bibitem{Mezard-Montanari-2009}
Mezard M and Montanari A 2009 {\em Information, Physics, and
Computation\/}
  (New York, USA: Oxford Univ. Press)

\bibitem{Mertens-etal-2006}
Mertens S, M{\'{e}}zard M and Zecchina R 2006 {\em Rand. Struct.
Algorithms\/}
  {\bf 28} 340--373

\bibitem{Montanari-etal-2008}
Montanari A, {Ricci-Tersenghi} F and Semerjian G 2008 {\em J. Stat.
Mech.:   Theor. Exp.\/}  P04004

\bibitem{Zhou-2008}
Zhou H 2008 {\em Phys. Rev. E\/} {\bf 77} 066102

\bibitem{Mezard-Parisi-2001}
M{\'{e}}zard M and Parisi G 2001 {\em Eur. Phys. J. B\/} {\bf 20}
217--233

\bibitem{Mezard-Montanari-2006}
M{\'{e}}zard M and Montanari A 2006 {\em J. Stat. Phys.\/} {\bf 124}
1317--1350

\bibitem{Semerjian-2008}
Semerjian G 2008 {\em J. Stat. Phys.\/} {\bf 130} 251--293

\bibitem{Ardelius-Zdeborova-2008}
Ardelius J and Zdeborova L 2008 {\em Phys. Rev. E\/} {\bf 78}
040101(R)

\bibitem{Gardner-1985}
Gardner E 1985 {\em Nucl. Phys. B\/} {\bf 257 [FS14]} 747--765

\bibitem{Zhou-2009-b}
Zhou H 2009 Criticality and heterogeneity in the solution space of
random constraint satisfaction problems arXiv:0911.4328 (Int. J. Mod. Phys. B,
accepted)

\bibitem{Zhou-Ma-2009}
Zhou H and Ma H 2009 {\em Phys. Rev. E\/} {\bf 80} 066108

\bibitem{Zhou-2009}
Zhou H 2010 {\em Eur. Phys. J. B\/} {\bf 73} 617--624

\bibitem{Li-Ma-Zhou-2009}
Li K, Ma H and Zhou H 2009 {\em Phys. Rev. E\/} {\bf 79} 031102

\bibitem{Wasserman-Faust-1994}
Wasserman S and Faust K 1994 {\em Social Network Analysis: Methods
and
  Applications\/} (New York: Cambridge University Press)

\bibitem{Girvan-Newman-2002}
Girvan M and Newman M~E~J 2002 {\em Proc. Natl. Acad. Sci. USA\/}
{\bf 99}
  7821--7826

\bibitem{Zhou-2003-b}
Zhou H 2003 {\em Phys. Rev. E\/} {\bf 67} 041908

\bibitem{Zhou-2003-c}
Zhou H 2003 {\em Phys. Rev. E\/} {\bf 67} 061901

\bibitem{Jain-Dubes-1988}
Jain A~K and Dubes R~C 1988 {\em Algorithms for Clustering Data\/}
(Englewood
  Cliffs, NJ, USA: Prentice-Hall)

\bibitem{Barthel-Hartmann-2004}
Barthel W and Hartmann A~K 2004 {\em Phys. Rev. E\/} {\bf 70} 066120

\bibitem{Zhou-2005b}
Zhou H 2005 {\em New J. Phys.\/} {\bf 7} 123

\bibitem{Cavagna-2009}
Cavagna A 2009 {\em Phys. Report\/} {\bf 476} 51--124

\bibitem{Ediger-2000}
Ediger M~D 2000 {\em Annu. Rev. Phys. Chem.\/} {\bf 51} 99--128

\bibitem{Glotzer-2000}
Glotzer S~C 2000 {\em J. Non-Cryst. Solids\/} {\bf 274} 342--355

\bibitem{note4}
A random $K$-SAT formula can be represented by a bipartite graph
(see, e.g.,
  Ref.~\cite{Kschischang-etal-2001}). In this bipartite graph, a circular node
  represents a variable and a square node represents a constraint, an edge
  between a variable node $i$ and a constraint node $a$ means that variable $i$
  is involved in constraint $a$. In this bipartite graph, the distance $d(i,j)$
  between two variable nodes $i$ and $j$ is defined as the number of constraint
  nodes on a shortest-length path between $i$ and $j$. For example, if $i$ and
  $j$ both are constrained by the same constraint, then $d(i,j)=1$; if $i$ and
  $j$ are not constrained by the same constraint but $i$ and a variable $k$ are
  involved in constraint $a$ and $j$ and $k$ are involved in constraint $b$,
  then $d(i,j)=2$.

\bibitem{Kschischang-etal-2001}
Kschischang F~R, Frey B~J and Loeliger H~A 2001 {\em IEEE Trans.
Infor.
  Theor.\/} {\bf 47} 498--519

\bibitem{Franz-Parisi-1995}
Franz S and Parisi G 1995 {\em J. de Physique I\/} {\bf 5} 1401--1415

\bibitem{Franz-Parisi-1997}
Franz S and Parisi G 1997 {\em Phys. Rev. Lett.\/} {\bf 79} 2486--2489

\bibitem{Zdeborova-Krzakala-2010}
Zdeborova L and Krzakala F 2010 {\em Phys. Rev. B\/} {\bf 81} 224202

\bibitem{Obuchi-Kabashima-2009}
Obuchi T and Kabashima Y 2009 {\em J. Stat. Mech.: Theor. Exp.\/} P12014


\end{thebibliography}
\providecommand{\newblock}{}

\end{document}